\documentclass[10pt, twocolumn]{article}
\pdfoutput=1
\usepackage{graphicx}

\begin{document}

\title{Biomimetic isotropic nanostructures for structural coloration}
\author{Jason D. Forster$^1$,
Heeso Noh$^2$,
Seng Fatt Liew$^2$,
Vinodkumar Saranathan$^3$\\
Carl F. Schreck$^4$,
Lin Yang$^5$,
Jin-Gyu Park$^1$,
Richard O. Prum$^3$, \\
Simon G. J. Mochrie$^{2,4}$,
Corey S. O'Hern$^{1,4}$,
Hui Cao$^{2,4}$,
and Eric R. Dufresne$^{1,4,6,7}$\\
\footnotesize{$^1$Mechanical Engineering, Yale University}\\
\footnotesize{$^2$Applied Physics, Yale University }\\
\footnotesize{$^3$Ecology and Evolutionary Biology, Peabody Museum of Natural History, Yale University}\\
\footnotesize{$^4$Physics, Yale University}\\
\footnotesize{$^5$National Synchrotron Light Source, Brookhaven National Laboratory}\\
\footnotesize{$^6$Chemical Engineering, Yale University}\\
\footnotesize{$^7$Cell Biology, Yale University}
}
\maketitle

Many species of birds have feathers that are brilliantly-colored without the use of pigments.
In these cases, light of specific wavelengths is selectively scattered from nanostructures with variations in index of refraction on length scales of the order of visible light. $^{\footnotesize{\cite{Srinivasarao:1999p9327}}}$
This phenomenon is called structural color.
The most striking examples of structural color in nature are iridescent colors created by scattering from periodic structures. $^{\footnotesize{\cite{Vukusic:2003p1380} \cite{Welch:2007p1313}}}$
The colors produced by these structures change dramatically depending on the angle of observation.
Nature also produces structural colors that have very little angle dependence.
These colors are the result of scattering from isotropic structures. $^{\footnotesize{\cite{Dufresne:2009p6342} \cite{Prum:2009p3119} \cite{Prum:1998p1228}}}$

In recent years, periodic biological structures have provided inspiration for groups trying to make photonic materials. $^{\footnotesize{\cite{Parker:2004p11036} \cite{Parker:2007p11434}}}$
Much of this work has been motivated by producing a photonic band gap. $^{\footnotesize{\cite{Yablonovitch:1987p1809} \cite{Joannopoulos:2008p11516} \cite{Noda:2003p11561} \cite{MSoukoulis:2001p11651} \cite{Kinoshita:2008p10935} \cite{Muller:2000p5865}}}$
However, Nature's alternative design, based on isotropic structures is just starting to be explored. $^{\footnotesize{\cite{Hallam:2009p11443} \cite{Takeoka:2009p10174}}}$
Hallam \emph{et al} have used biomimetic random structures to make ultra-thin mineral coatings that are brilliant white.
Takeoka \emph{et al} recently showed  that a wide range of colors with very little angle dependence can be produced by microgel dispersions.
In this communication, we describe the self-assembly of biomimetic isotropic films which display structural color that is amenable to  potential applications in coatings, cosmetics, and textiles.
We find that isotropic structures with a characteristic length-scale comparable to the wavelength of visible light can produce structural color when wavelength-independent scattering is suppressed.

\begin{figure}
	\begin{center}
	\includegraphics[width=3in]{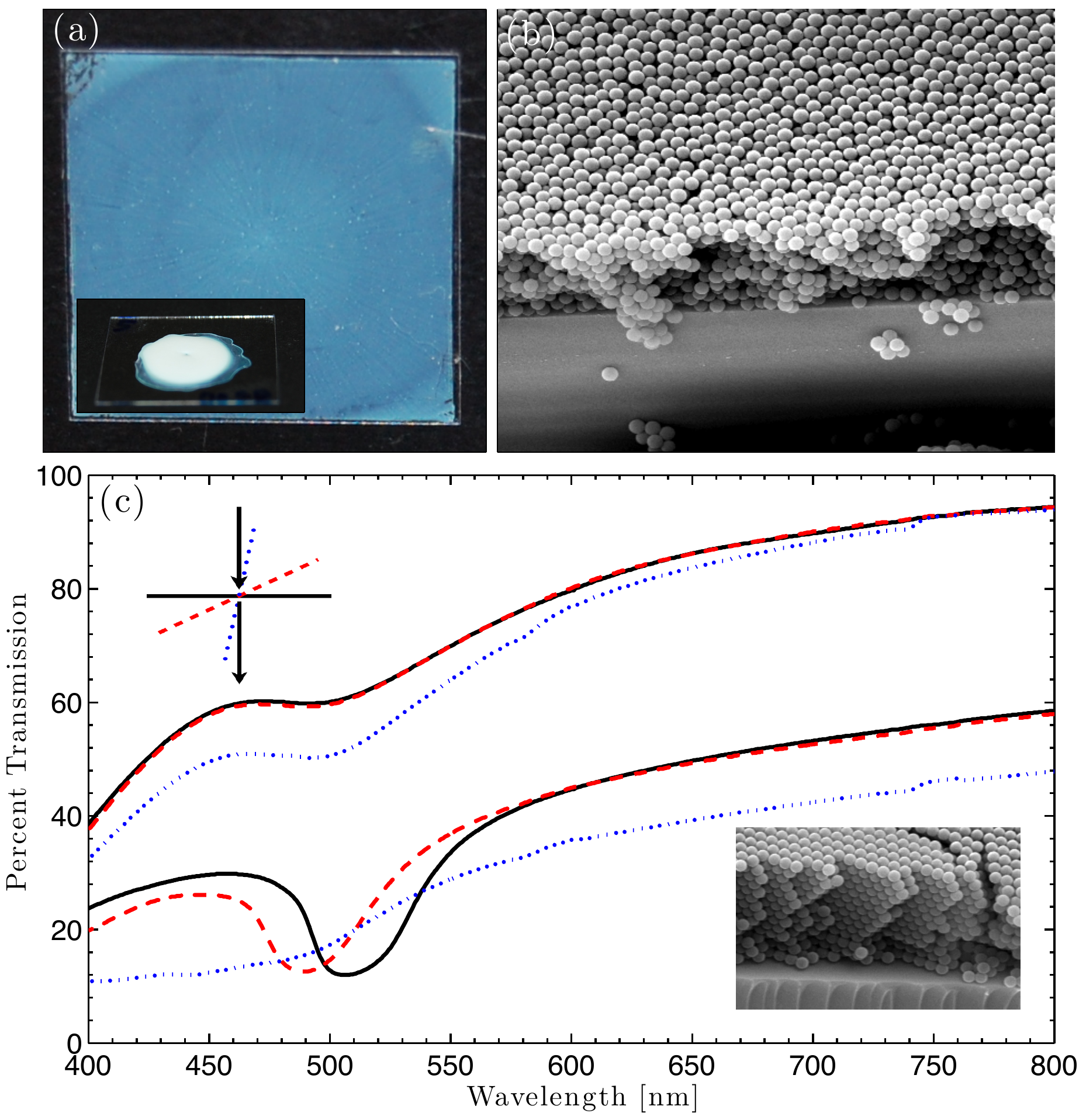}
	\end{center}
	\caption{
		\emph{Effect of disorder and order on optical properties.}
		\emph{a)} Photograph of a film of 226 and 271 nm PS spheres spin-cast onto an 18 $\times$ 18 mm glass coverslip.
		\emph{Inset}: Photograph of a dried sessile droplet.
		\emph{b)} Side-view SEM image a similar film. The field of view is 8.8 $\mu$m wide.
		\emph{c)} The top set of curves show the transmission spectra for the isotropic film pictured in (a).
		The bottom set of curves show the results of the same measurement performed with a crystalline sample, made by spin-casting 226 nm PS spheres.
		The data were taken with the sample normal to the optical path (solid line), at an angle of 30$^{\circ}$ (dashed red line), and at an angle of 80$^{\circ}$ (dotted blue line), represented schematically in the upper left hand corner.
		\emph{Inset}:  Side-view SEM image of a crystalline sample, the field of view is 5.3 $\mu$m wide.}
	\label{transmission}
\end{figure}

We make two types of films that are structurally-colored by exploiting the self-assembly of colloidal polymer nanoparticles.
The first type of sample is a thin film on a glass coverslip produced by spin casting an aqueous suspension of spheres (Figure~\ref{transmission}a).
While monodisperse dispersions form anistropic polycrystalline films, as shown in the inset of Figure~\ref{transmission}c, a mixture of two sizes of spheres ensures an isotropic structure as shown in Figure~\ref{transmission}b.$^{\footnotesize{\cite{Sear:1998p4095}}}$
The blue-green film  in Figure~\ref{transmission} is made from a bidisperse suspension of polystyrene (PS) spheres with mean diameters of 226 and 271 nm and polydispersity 2\% in equal volume fractions.

We image the structure of the film using scanning electron microscopy (SEM).
The side-view SEM in Figure~\ref{transmission}b is of a film comparable to the one in Figure~\ref{transmission}a and shows that there is no long range order, it also reveals a representative thickness of 2.3$\pm$0.2 $\mu$m.

The optical properties of isotropic and crystalline films are quite different, even for samples prepared with very similar particles.
We compare the transmission spectra of isotropic and crystalline samples in Figure~\ref{transmission}c.
The isotropic film is the same one pictured in Figure~\ref{transmission}a, the crystalline film is composed solely of the 226 nm spheres.
Both samples show a dip near 500 nm at normal incidence.
However, when the angle of incidence is changed, the spectral position of the dip for the crystalline sample shifts and eventually disappears.
In contrast,  the position of the dip for the isotropic sample does not move.
This illustrates the trade-off in optical performance between crystalline and isotropic structures: while the crystalline film has more pronounced features at some angles, the isotropic film performs consistently over a wide range of angles.

\begin{figure}
	\begin{center}
	\includegraphics[width=3in]{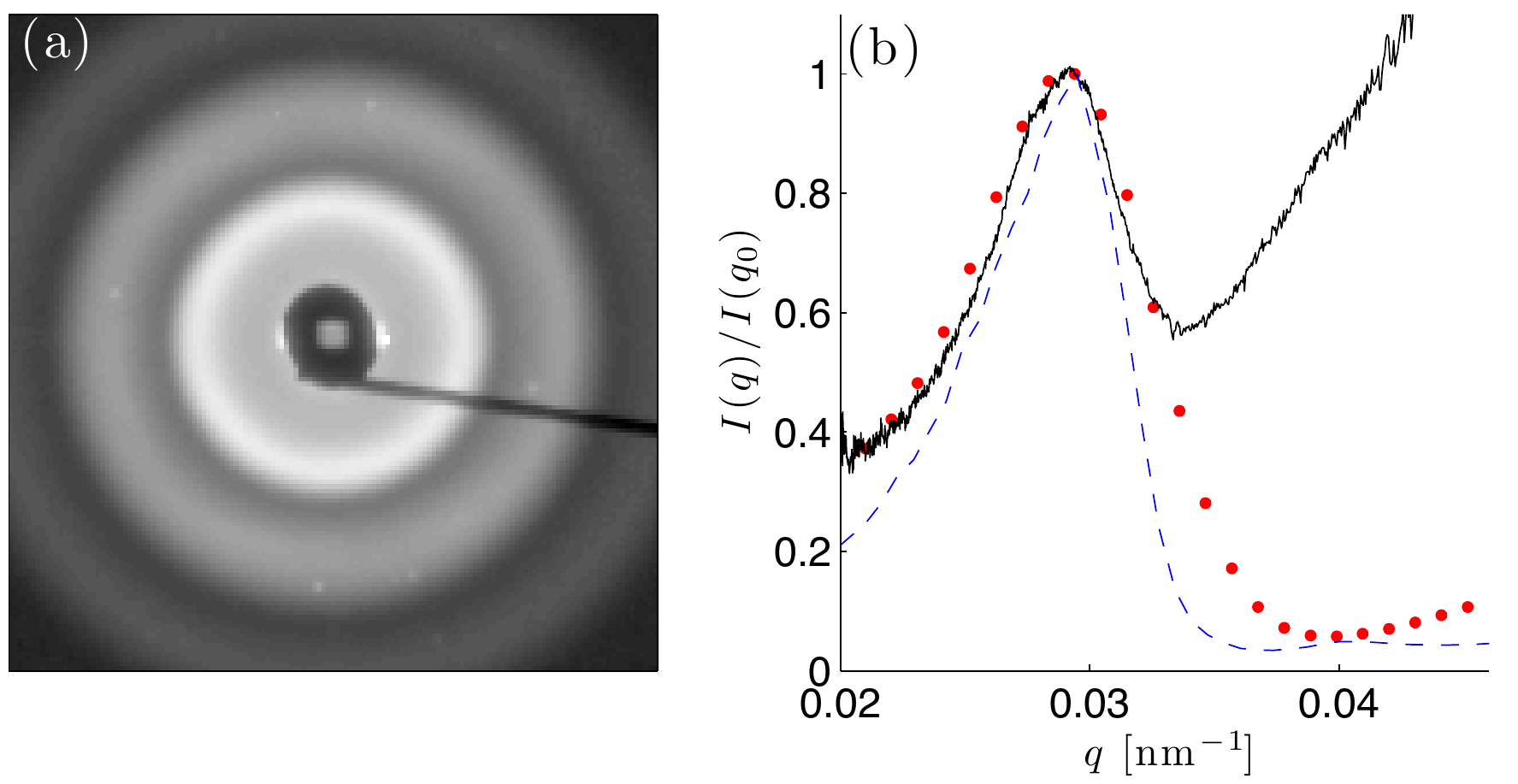}
	\end{center}
	\caption{
		\emph{Structure of isotropic films}.
		\emph{a)} SAXS pattern from an isotropic film, the field of view is 0.15 nm$^{-1}$ and the gray values are logarithmic in intensity.
		 \emph{b)} The azimuthal average of the experimental (red dots) and numerical (blue dashed line) scattering pattern.  The black line is optical reflectivity data taken at an angle of 10$^{\circ}$ (experiment geometry shown in Figure~\ref{qseries}) converted to q-space using an effective refractive index of $n_{e}=1.24$.}
	\label{structure}
\end{figure}

To quantify the structure of the films, we perform small-angle X-ray scattering (SAXS) measurements on samples composed of spheres with mean diameters of 226 and 265 nm with 2\% polydispersity, spin-cast on kapton tape.
A typical scattering pattern is shown in Figure~\ref{structure}a.
The pattern is dominated by a ring of uniform intensity, indicating that there is a well-defined length scale with no preferred direction \--- the structure is isotropic with short-range order.
The azimuthal average of this pattern is shown in Figure~\ref{structure}b, along with the expected scattering intensity from simulations of jammed packings of bidisperse spheres with size and number ratios that match our experimental system.
The position of the first peak in $I(q)$ occurs at $q_0=$ 0.03 nm$^{-1}$ for both simulation and experiment.
The full width at half maximum (FWHM) of the first peak, $\Delta q$, characterizes the range of spatial order $\xi = 2\pi/\Delta q$ = 870 nm.
In powder crystallography, $\xi$ describes the crystal domain size, here, $\xi$ is only a few particle diameters.

We directly compare optical reflectivity measurements to SAXS measurements by plotting the reflected intensity as a function of wavevector $q = (4\pi n_{e}/\lambda)\cos{(\theta_m/2)}$, where $n_{e}$ is the effective refractive index of the material;  $\theta_m$ is the angle between illumination and detection, taking into account refraction at the film surface; and $\lambda$ is the wavelength of light in vacuum (Figure~\ref{structure}b).
The peaks from both measurements match when $n_{e} =$ 1.24.  We apply the Maxwell-Garnett equation,
$$n_{e} = n_{air}\left(\frac{2n_{air}^2 + n_{PS}^2 + 2\phi(n_{PS}^2 - n_{air}^2)}{2n_{air}^2 + n_{PS}^2 - \phi(n_{PS}^2 - n_{air}^2)}\right)^{1/2}$$
to calculate the volume fraction of spheres, $\phi =$ 0.46$\pm$0.04, where $n_{PS}$ is the index of refraction of the spheres, taken to be 1.58, and $n_{air}$ is taken to be 1.00.

Isotropic films can produce structural color with little angle dependence, but the film thickness  critically affects its color, as seen in the inset of Figure~\ref{transmission}a.
Here, we cast a thick film by  drying  a sessile droplet of the same suspension used to make the thin films.
In thick regions near the center, the film appears white.
In thin sections near the edge, it appears blue-green.
This thickness dependence can be understood in the following way.
The film preferentially scatters wavelengths corresponding to the peak in $I(q)$.
In a thin film, only these wavelengths will be scattered to the detector resulting in a structural color.
In a thick film, all wavelengths are scattered multiple times and reach the detector.

The sensitivity of the color to the film thickness requires well-controlled casting procedures which increase the cost and limit the coated area.
Therefore, for many potential applications it is necessary to eliminate the thickness-dependence.
We address this by introducing broadband absorption to the bulk of the films.
The absorption length plays a similar role as the thickness: it limits the path length of light through the film by absorbing photons that do not get scattered within a small distance from the surface.
Thus, only wavelengths with the strongest scattering will escape the film before being absorbed.

\begin{figure}
	\begin{center}
	\includegraphics[width=3in]{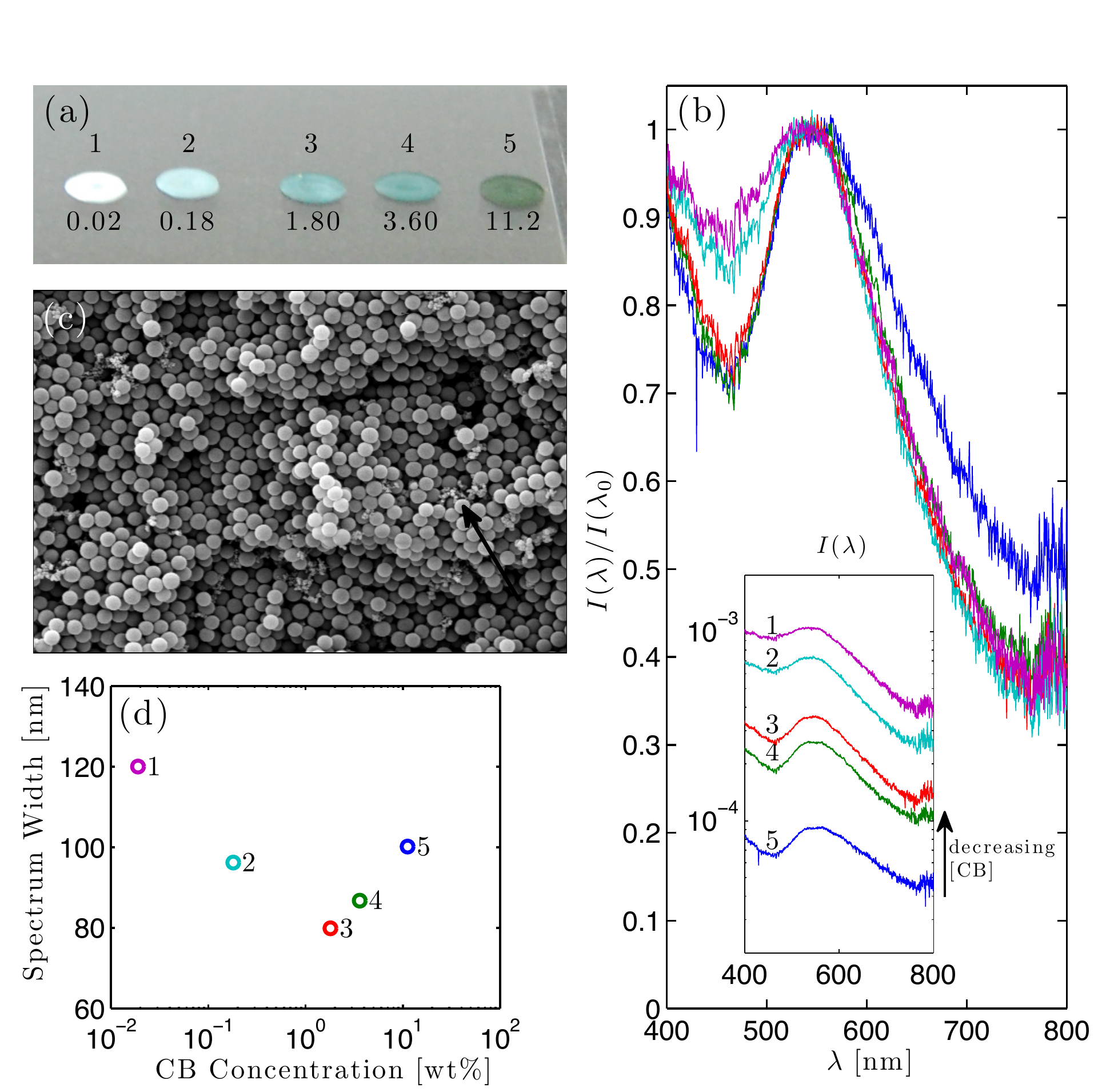}
	\end{center}
	\caption{
		\emph{Optimizing color by adding absorption.}
		\emph{a)} Photograph of five drop-cast films of 226 and 265 nm PS spheres containing carbon black.  Sample numbers and [CB] in wt$\%$ appear above and below the samples, respectively.
		\emph{b)} Normalized optical scattering spectra recorded at an angle of 20$^{\circ}$ (experiment geometry shown in Figure~\ref{qseries}) for the five samples in (a).
		\emph{Inset}: Non-normalized scattering spectra for the same samples.
		\emph{c)} SEM image of the interior region of sample 3, the field of view is 7.8 $\mu$m wide, a piece of CB is indicated with an arrow.
		\emph{d)} The width of the spectra for each sample in (a) at $I(\lambda)/I(\lambda_0) = 0.90$}
	\label{carbonblack}
\end{figure}

We make a series of thick films with different absorption lengths by drop-casting films with varying concentrations of carbon black.
The photo in Figure~\ref{carbonblack}a illustrates the effect of adding carbon black on the color of the material.
From left to right, the concentration of carbon black, [CB], is increased.
Intuitively, one might expect a mixture of black and white to make gray, but here they make blue-green over a range of [CB].
The plot in Figure~\ref{carbonblack}b shows the normalized optical scattering spectra for these five samples.
As [CB] is increased from 0.02 wt\% to 1.80 wt\%, the contrast between scattered intensity at the peak and shorter wavelengths is increased, improving the color of the sample.
When [CB] = 11.2 wt\%, the scattered intensity is reduced by a factor of ten compared to the brightest sample (inset in Figure~\ref{carbonblack}b), and the film appears dark gray.
The reflectance peak is narrowest for [CB] = 1.80 wt\% (Figure~\ref{carbonblack}d), and indeed this sample has a distinct blue-green color.
Table~\ref{carbonblacktable} lists the [CB] for each sample and the corresponding extinction length: the extinction length is extrapolated from measurements of aqueous suspensions of CB.
Interestingly, the extinction length for sample 3 is 1.3 $\mu$m, which is comparable to the thickness of the thin film pictured in Figure~\ref{transmission}a.

\begin{table}[htdp]\footnotesize
\caption{Carbon Black Extinction Length}
\begin{center}
\begin{tabular}{c c c}
Sample & [CB] [wt\%] & Extinction Length [m] \\
\hline
1 & 0.02 & 1.3$\times$10$^{-4}$\\
2 & 0.18 & 1.3$\times$10$^{-5}$\\
3 & 1.80 & 1.3$\times$10$^{-6}$\\
4 & 3.60 & 6.5$\times$10$^{-7}$\\
5 & 11.2 & 2.1$\times$10$^{-7}$\\
\end{tabular}
\end{center}
\label{carbonblacktable}
\end{table}

\begin{figure}
	\begin{center}
	\includegraphics[width=3in]{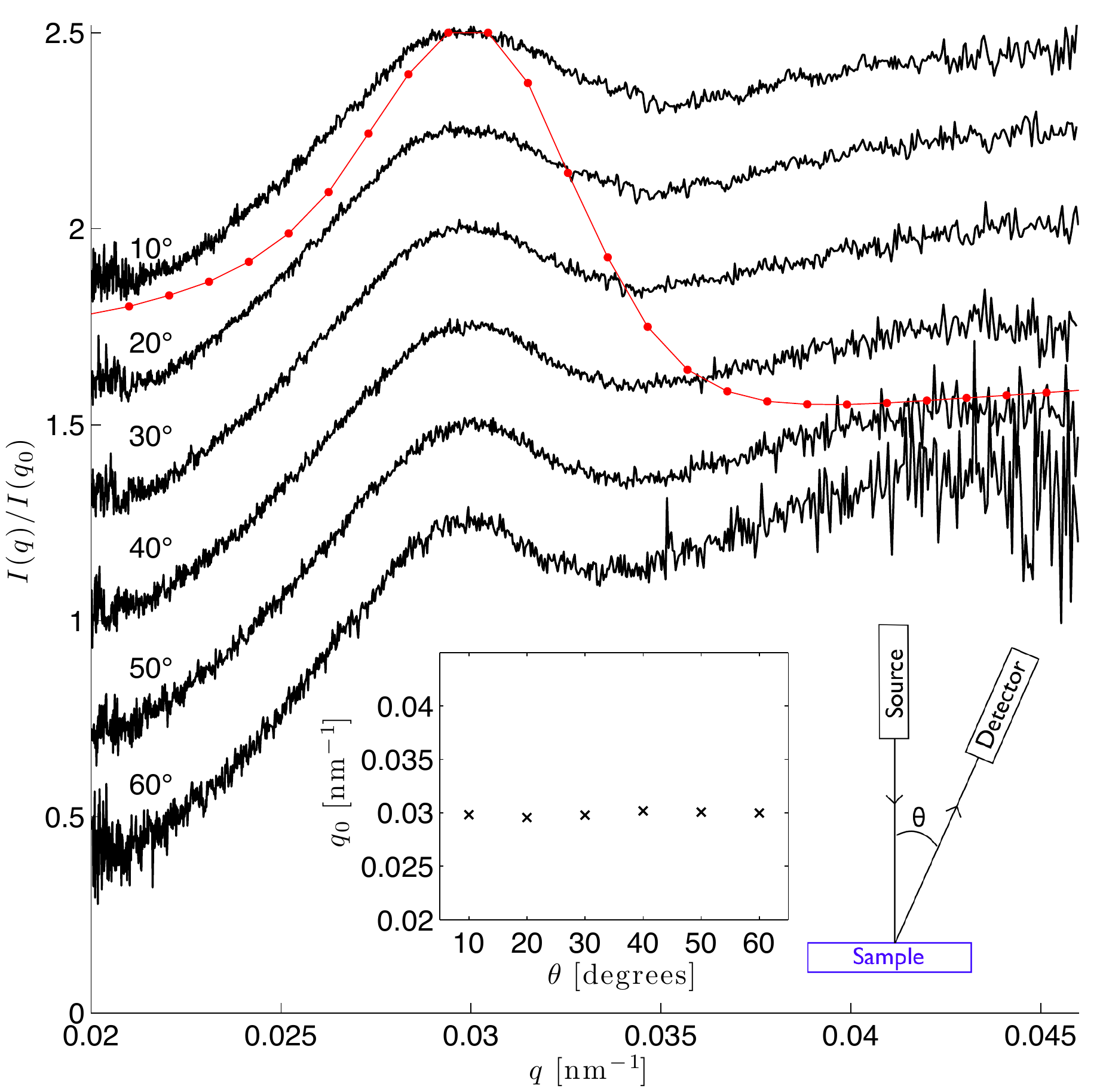}
	\end{center}
	\caption{
		\emph{Color is the result of single scattering.}
		Normalized reflectance spectra taken over a range of angles and converted to q-space for sample 2 in Figure~\ref{carbonblack} (black lines), and the azimuthal average of the SAXS pattern for a similar sample (red dots).
		The spectra have been vertically offset for clarity.
		\emph{Inset}: Position of the peak as a function of angle.
		\emph{Schematic}: the source and sample are fixed and the detector is rotated.}
	\label{qseries}
\end{figure}

The thick films with CB are not iridescent under omnidirectional illumination, but, under directional illumination, the peak wavelength scattered does change slightly when the angle between illumination and detection is varied. $^{\footnotesize{\cite{Noh:2009AM}}}$
Since the colors are the result of single scattering, the position of the scattering intensity maximum does not vary with respect to $q$, as demonstrated by the spectra in Figure~\ref{qseries}.
The red dots connected by a red line in Figure~\ref{qseries} represent the azimuthal average of the SAXS pattern for a sample with the same [CB] prepared on kapton tape.  The peaks from the optical and SAXS measurements match when $n_{e} = $ 1.29, implying that $\phi_{s}$ = 0.54$\pm$0.05.
The higher value of $\phi_{s}$ for the thick films relative to the thin films is most likely the result of the different quench-rates used to make the samples.
In the spin casting of thin films, the water evaporates within seconds, while water evaporates from the thick films over a few hours, allowing particle rearrangement which leads to a higher $\phi_s$.
In the thick film, we find that $\xi = $ 940 nm, only a few particle diameters.

\begin{figure}
	\begin{center}
	\includegraphics[width=3in]{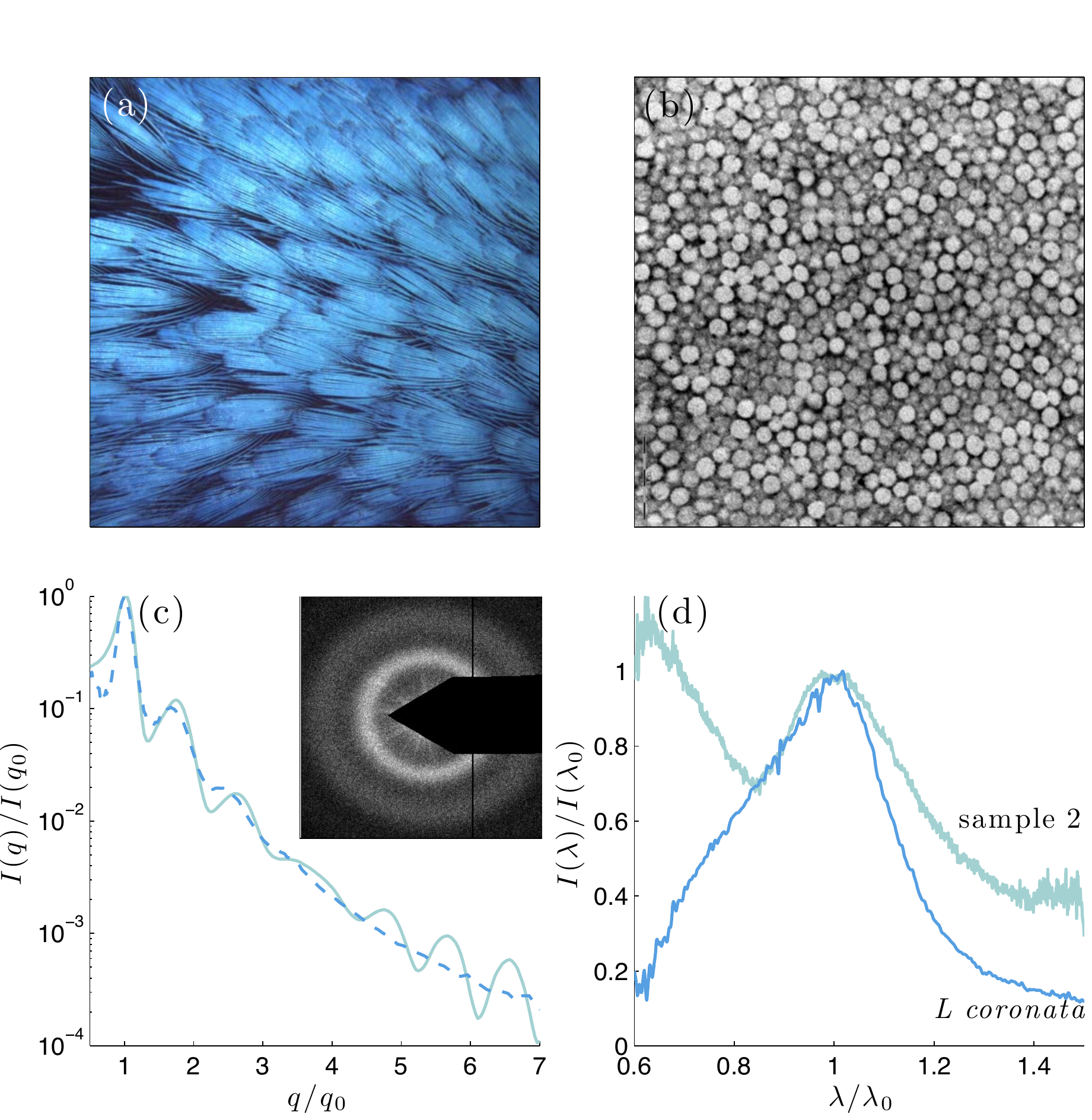}
	\end{center}
	\caption{
		\emph{Comparison with feathers.}
		\emph{a)} Photograph of crown feathers from \emph{L. coronata}, the field of view is 1 cm.
		 \emph{b)} TEM image of air spheres in beta-keratin from \emph{L. coronata}, the field of view is 5.5 $\mu$m.
		 \emph{c)} Azimuthal averages from SAXS measurements of \emph{L. coronata} (dashed line) and a thick film with the same [CB] as sample 2 in Figure~\ref{carbonblack} (solid line).
		 \emph{Inset:} SAXS pattern from \emph{L. coronata}, the field of view is 0.14 nm$^{-1}$ and the gray values are logarithmic in intensity.
		 \emph{d)} Comparison of optical scattering data taken at an angle of 20$^{\circ}$ (experiment geometry shown in Figure~\ref{qseries}) from \emph{L. coronata} and sample 2 from Figure~\ref{carbonblack}.  The line colors correspond to the apparent color of the feathers and the thick film.}
	\label{birdcomparison}
\end{figure}

These isotropic films mimic the essential optical properties of bird feathers that have structural color from isotropic nanostructures.
A photograph of an example of these feathers from the crown of \emph{Lepidothrix coronata} is shown in Figure~\ref{birdcomparison}a, and a transmission electron micrograph (TEM) of the color-producing structure is shown in Figure~\ref{birdcomparison}b.
Normalized $I(q)$ from both \emph{L. coronata} and sample 2 from Figure~\ref{carbonblack} are plotted as a function of $q/q_{0}$ in Figure~\ref{birdcomparison}c.
The SAXS patterns reveal similar structures out to the third peak in $I(q)$.
Beyond that, the thick film has additional peaks that arise from the uniformity of the spheres used in the sample.
We compare the performance of the feathers and films by plotting optical scattering spectra for both at 20$^\circ$ in Figure~\ref{birdcomparison}d.
The scattered optical intensity peak for \emph{L. coronata} is narrower than the film: the full-width at $I(\lambda)/I(\lambda_0) = 0.90$ is 49 nm for the feather and 82 nm for the film.
Similarly, \emph{L. coronata} has a narrower first peak in $I(q)$.
\emph{L. coronata} also displays less scattering at shorter wavelengths.
This may be due to a significant difference in the two structures: the feathers have spheres of air in a high-index of refraction background whereas the films have spheres of a high-index in a background of air.

Non-iridescent structural color is realized when an isotropic structure has a peak in $I(q)$ and wavelength-independent scattering is suppressed.
Using a bidisperse mixture of spheres produces isotropic structures.
Controlling film thickness or absorption length are effective at enhancing contrast in reflectance spectra.
These films may have applications in coatings, cosmetics and textiles.
The basic optical properties of structurally-colored feathers have been reproduced, but more work must be done before the biomimetic samples perform as well as their biological counterparts.

The authors would like to acknowledge seed funding from the Yale NSF MRSEC (DMR-0520495), a NSF CAREER grant to ERD (CBET-0547294), and a NSF grant to HC (DMR-0808937).
Research carried out in part at the National Synchrotron Light Source (NSLS), Brookhaven National Laboratory, which is supported by the U.S. Department of Energy.
Research carried out in part at the Advanced Photon Source (APS) at Argonne National Labs with the help of Drs. Alec Sandy and Suresh Narayanan, and supported by the U.S. Department of Energy, Office of Science, Office of Basic Energy Sciences, under Contract No. DE-AC02-06CH11357.

\section{Experimental}
Monodisperse PS spheres were synthesized using a surfactant-free  polymerization technique. \cite{CHONDE:1981p1047}
Particles were electrostatically stabilized by co-polymerization with sodium 4-vinylbenzenesulfonate.
After synthesis, particle suspensions were washed by centrifugation and resuspension at least three times with DI water.
After washing, all particle suspensions were adjusted to $\phi_{PS}$ = 0.3.
Particle sizes were determined by SEM image analysis using a Philips XL-30 ESEM with an accelerating voltage of 10 kV, after being coated with a thin layer of gold.

To prepare bidisperse suspensions, equal volumes of two monodisperse suspensions were mixed by pipetting approximately 20 times and vortexing for at least 30 seconds.
Prior to spin casting for thin films or water evaporation for thick films, the suspensions were sonicated for at least 20 minutes.
Thin films were spin cast onto glass coverslips cleaned with ethanol using a Headway Research, Inc. PWM32-PS-R790 spin coater.
Typical spin speeds were between 500 and 5000 RPM, the spin speed determined the final thickness of the film. \cite{Jiang:2004p8684}

Cabot Vulcan XC72R GP-3919 carbon black was suspended in DI water at [CB] = 4.2 wt$\%$ with 1 wt$\%$ Pluronic F108 to stabilize the CB.
Different volumes of the CB suspension were added to bidisperse suspensions of spheres to produce samples with different final [CB].

To estimate the extinction lengths quoted in Table~\ref{carbonblacktable}, the transmission spectra of aqueous suspensions of CB were measured using an Ocean Optics USB 650 Red Tide spectrometer.
Suspensions ranging from 8$\times$10$^{-5}$ to 2$\times$10$^{-2}$ wt$\%$ were used.
Measurements were made with three different path lengths: 1 cm, 0.04 cm, and 0.03 cm.
The extinction length for all path lengths was proportional to [CB]$^{-1}$, allowing us to extrapolate the extinction length to the [CB] range used in our thick film samples.
We use extinction instead of absorption because our measurement did not discriminate between absorbed and scattered light.

SAXS measurements of biomimetic samples were carried out at beamline X9 at the NSLS, Brookhaven Naitonal Laboratory, using a Rayonix Mar 165 CCD detector.
The X-ray energy was 7 keV (a wavelength of 1.771 \AA) and the sample-to-detector distance was ~5 m.
The conversion from the detector image to reciprocal space was calibrated using the diffraction pattern from a standard silver behenate sample ($q_0=$ 0.1076 \AA$^{-1}$).
The first ring from silver behenate is out of the angular range covered by the detector at 7 keV, therefore the standard pattern was collected with the same scattering geometry but at an X-ray energy of 15.65 keV (0.792 \AA).
The finite beam size at the detector corresponds to FWHM $q$ resolution of 0.00045 \AA$^{-1}$, which is $\sim$17$\%$ of the width of the fringe produced by the spheres ($\pi/120$ nm $\approx$ 0.0026 \AA$^{-1}$).
Biomimetic samples for SAXS measurements were prepared on 0.0025"-thick Kapton Tape, purchased from McMaster-Carr (catalog no. 7648A33).

SAXS measurements of the feathers were carried out at beamline 8-ID-I at the APS, Argonne National Labs, as described in Dufresne \emph{et al}. \cite{Dufresne:2009p6342}

To simulate the structure of our samples, we created mechanically stable packings of bidisperse frictionless spheres in cubic cells with periodic boundary conditions using a simulation protocol in which we successively compress/decompress soft particles and then apply conjugate gradient energy minimization until particles are just at contact (up to a prescribed energy threshold). \cite{Gao:2006p061304}
The energy threshold, initial packing fraction $\phi_0$, and increment in $\phi$ were set to $10^{-8}$, $0.2$ and $10^{-4}$, respectively.
We selected particle size and number ratios to match the experiments.
From $50$ independent initial configurations, we obtained an average packing fraction $\phi_J = 0.63$, which is relatively insensitive to the specific parameters of the packing-generation protocol.
From the particle centers, we calculated the partial, $S_{ll}(q)$, $S_{ss}(q)$, and $S_{ls}(q)$, and total structure factor $S(q) = x_l S_{ll}(q) + x_s S_{ss}(q) + 2\sqrt{x_l x_s} S_{ls}(q)$, where $l,s$ signify large or small particles, $x$ is the number fraction of the indicated particle, and $q$ is the wavevector.
The partial structure factors and the theoretical form factors for each sphere size were used to calculate the total scattering intensity $I(q) \propto x_l S_{ll}(q) F_l^2(q) + 2\sqrt{x_l x_s}S_{ls}(q) F_l(q) F_s(q) + x_s S_{ss}(q) F_s^2(q)$.

Optical transmission spectra were performed using a Hitachi U-2001 spectrophotometer.
Optical reflection and scattering spectra were performed with a custom-built setup in Hui Cao's laboratory described in detail in Noh \emph{et al}. \cite{Noh:2009AM}

\bibliographystyle{ieeetr}

\begin{thebibliography}{10}

\bibitem{Srinivasarao:1999p9327}
M.~Srinivasarao, ``Nano-optics in the biological world: beetles, butterflies,
  birds, and moths,'' {\em Chemical reviews}, vol.~99, no.~7, pp.~1935--1961,
  1999.

\bibitem{Vukusic:2003p1380}
P.~Vukusic and J.~R. Sambles, ``Photonic structures in biology,'' {\em Nature},
  vol.~424, 2003.

\bibitem{Welch:2007p1313}
V.~L. Welch and J.~P. Vigneron, ``Beyond butterflies---the diversity of
  biological photonic crystals,'' {\em Optical and Quantum Electronics},
  vol.~39, pp.~295--303, 2007.

\bibitem{Dufresne:2009p6342}
E.~R. Dufresne, H.~Noh, V.~Saranathan, S.~G.~J. Mochrie, H.~Cao, and R.~O.
  Prum, ``Self-assembly of amorphous biophotonic nanostructures by phase
  separation,'' {\em Soft Matter}, vol.~5, no.~9, pp.~1792--1795, 2009.

\bibitem{Prum:2009p3119}
R.~O. Prum, E.~R. Dufresne, T.~Quinn, and K.~Waters, ``Development of
  colour-producing beta-keratin nanostructures in avian feather barbs,'' {\em
  Journal of The Royal Society Interface}, vol.~6, pp.~S253--S265, 2009.

\bibitem{Prum:1998p1228}
R.~O. Prum, R.~H. Torres, S.~Williamson, and J.~Dyck, ``Coherent light
  scattering by blue feather barbs,'' {\em Nature}, vol.~396, pp.~28--29, 1998.

\bibitem{Parker:2004p11036}
A.~R. Parker, ``A vision for natural photonics,'' {\em Philosophical
  Transactions of the Royal Society A: Mathematical, Physical and Engineering
  Sciences}, vol.~362, no.~1825, pp.~2709--2720, 2004.

\bibitem{Parker:2007p11434}
A.~R. Parker and H.~E. Townley, ``Biomimetics of photonic nanostructures,''
  {\em Nature Nanotechnology}, vol.~2, pp.~347--353, 2007.

\bibitem{Yablonovitch:1987p1809}
E.~Yablonovitch, ``Inhibited spontaneous emission in solid-state physics and
  electronics,'' {\em Phys. Rev. Lett.}, vol.~58, no.~20, pp.~2059--2062, 1987.

\bibitem{Joannopoulos:2008p11516}
J.~D. Joannopoulos, R.~D. Meade, and J.~N. Winn, {\em Photonic Crystals:
  Molding the Flow of Light}.
\newblock Princeton University Press, 1995.

\bibitem{Noda:2003p11561}
S.~Noda and T.~Baba, {\em Roadmap on Photonic Crystals}.
\newblock Kluwer Academic Publishers, 2003.

\bibitem{MSoukoulis:2001p11651}
C.~M. Soukoulis, {\em Photonic Crystals and Light Localization in the 21st
  Century}.
\newblock Springer, 2001.

\bibitem{Kinoshita:2008p10935}
S.~Kinoshita, S.~Yoshioka, and J.~Miyazaki, ``Physics of structural colors,''
  {\em Reports on Progress in Physics}, vol.~71, no.~7, p.~076401 (30pp), 2008.

\bibitem{Muller:2000p5865}
M.~Muller, R.~Zentel, T.~Maka, S.~G. Romanov, and C.~M.~S. Torres, ``Photonic
  crystal films with high refractive index contrast,'' {\em Advanced
  Materials}, vol.~12, no.~20, 2000.

\bibitem{Hallam:2009p11443}
B.~T. Hallam, A.~G. Hiorns, and P.~Vukusic, ``Developing optical efficiency
  through optimized coating structure: biomimetic inspiration from white
  beetles,'' {\em Applied Optics}, vol.~48, no.~17, pp.~3243--3249, 2009.

\bibitem{Takeoka:2009p10174}
Y.~Takeoka, M.~Honda, T.~Seki, M.~Ishii, and H.~Nakamura, ``Structural colored
  liquid membrane without angle dependence,'' {\em ACS Appl. Mater.
  Interfaces}, vol.~1, no.~5, pp.~982--986, 2009.

\bibitem{Sear:1998p4095}
R.~P. Sear, ``Phase separation and crystallisation of polydisperse hard
  spheres,'' {\em Europhysics Letters}, vol.~44, no.~4, pp.~531--535, 1998.

\bibitem{Noh:2009AM}
H.~Noh, S.~F. Liew, V.~Saranathan, R.~O. Prum, S.~G.~J. Mochrie, E.~R.
  Dufresne, and H.~Cao, ``How non-iridescent colors are generated by
  quasi-ordered structures of bird feathers,'' {\em Advanced Materials},
  preprint.

\bibitem{CHONDE:1981p1047}
Y.~Chonde and I.~M. Krieger, ``Emulsion polymerization of styrene with ionic
  comonomer in the presence of methanol,'' {\em Journal of Applied Polymer
  Science}, vol.~26, no.~6, pp.~1819--1827, 1981.

\bibitem{Jiang:2004p8684}
P.~Jiang and M.~J. McFarland, ``Large-scale fabrication of wafer-size colloidal
  crystals, macroporous polymers and nanocomposites by spin-coating,'' {\em J.
  Am. Chem. Soc}, vol.~126, pp.~13778--13786, 2004.

\bibitem{Gao:2006p061304}
G.-J. Gao, J.~B\l{}awzdziewicz, and C.~S. O'Hern, ``Frequency distribution of
  mechanically stable disk packings,'' {\em Phys. Rev. E}, vol.~74, p.~061304,
  Dec 2006.

\end{thebibliography}

\end{document}